\documentclass[a4paper, 10pt,conference]{ieeeconf}

\usepackage[utf8]{inputenc}

\usepackage{amsmath,amssymb}
\usepackage{graphicx}
\usepackage{xspace}
\usepackage[dvipsnames]{xcolor}
\usepackage{stmaryrd}
\usepackage{subcaption}
\usepackage{array}
\usepackage{hyperref}
\usepackage{multirow}
\usepackage{geometry}
\begin{document}

\title{Automatic Driver Identification from In-Vehicle Network Logs}

\author{
\authorblockN{Mina Remeli, Szilvia Lesty{\'a}n, Gergely Acs, and Gergely Bicz{\'o}k}
\authorblockA{CrySyS Lab, BME-HIT, Hungary\\
\{remeli, lestyan, acs, biczok\}@crysys.hu}}

\maketitle

\begin{abstract}
Data generated by cars is growing  at an  unprecedented scale. As cars gradually become part of the Internet of Things (IoT) ecosystem, several stakeholders discover the value of in-vehicle network logs containing the measurements of the multitude of sensors deployed within the car. This wealth of data is also expected to be exploitable by third parties for the purpose of profiling drivers in order to provide personalized, value-added services. Although several prior works have successfully demonstrated the  feasibility  of  driver  re-identification  using  the  in-vehicle  network  data  captured  on the  vehicle’s  CAN (Controller Area Network) bus, they inferred the identity of the driver only from known sensor signals (such as the vehicle's speed, brake pedal position, steering wheel angle, etc.) extracted from the CAN messages. However, car manufacturers intentionally do not reveal exact signal location and semantics within CAN logs. We show that the inference of driver identity is possible even with off-the-shelf machine learning techniques without reverse-engineering the CAN protocol. We demonstrate our approach on a dataset of 33 drivers and show that a driver can be re-identified and distinguished from other drivers with an accuracy of 75-85\%.
\end{abstract}
\section{Introduction}
\label{sec:intro}

Almost all vehicles in use nowadays are equipped with various on-board Electrical Control Units (ECUs), sensors and actuators measuring and controlling the vehicle's speed, acceleration, braking, fuel consumption, battery status, or tire pressure level, among others.
 Sensors attached to their respective ECUs, whose number ranges from 5 to a hundred per vehicle, continuously generate a vast amount of real-time data. In order to implement complex control tasks for traffic safety or passenger infotainment, these data are then transferred among ECUs and other nodes over the in-vehicle network, most commonly following the Controller Area Network (CAN bus) standard~\cite{szalay2015ict}. In addition to being used real-time for automotive control, data are also 
 transferred to car manufacturers through Internet-connected devices or pre-installed modems every half to five minutes.  Apart from diagnostic and forensics purposes of car companies and their primary data analytics partners, these huge datasets also become exploitable by other parties such as insurance companies, retailers, advertisers and parking operators; in general, vehicular data gradually become part of the Internet of Things (IoT) ecosystem\footnote{\href{https://www.forbes.com/sites/forbestechcouncil/2018/05/08/how-well-does-your-car-know-you/}{www.forbes.com/sites/forbestechcouncil/2018/05/08/how-well-does-your-car-know-you/}}.
  Moreover, upcoming autonomous cars produce and may share an order of magnitude larger amount of real-time data ($\approx$ 1~GB/s) than traditional cars.
 Indeed, data from vehicles are predicted to create a \$10 trillion market and could become five times bigger than the market for the cars themselves in the next few years\footnote{\href{https://bit.ly/2V14YkK}{bit.ly/2V14YkK}}.
 
 Unfortunately, sharing in-vehicle network data raises serious privacy concerns. Although drivers are expected to opt-in to such data sharing\footnote{\href{https://www.wsj.com/articles/what-your-car-knows-about-you-1534564861}{www.wsj.com/articles/what-your-car-knows-about-you-1534564861}}, it is still unclear what exact personal information they would transfer then to third parties.
 For example, can a skilled data analyst infer the driver's identity using \emph{only} in-vehicle network data?   
 Despite the inherently noisy nature of this fine-grained measurement data, the feasibility of driver identification has been demonstrated in several prior works~\cite{EnevTKK16, HallacSSLHRSL16, GMM_driving, Zhang:2016:DCB:2856767.2856806}.
 In particular, it is well-known that drivers can be re-identified in constrained environments if they follow the same route with the same car and sensor readings are available from the captured network logs~\cite{EnevTKK16}. However, when following different routes, unique driving patterns are more difficult to extract due to the variable traffic conditions. Also, it is much more plausible that an adversary is able to collect CAN logs from arbitrary routes, hence we stick with this scenario throughout this paper. Moreover, car manufacturers do not disclose the exact format of CAN messages in order to protect their intellectual properties against competitors, as well as the security of drivers against malicious car hacking. But is this approach of security/privacy by obscurity  effective? 

A straightforward solution could be to reverse-engineer the CAN protocol by extracting different sensor signals, such as velocity, steering wheel angle, etc., from the captured CAN messages. However, this requires the manual inspection of every possible time series whether they contain a given signal with sufficient predictive power. This process is tedious and often considered to be illegal \cite{bowers_legal}.

In this paper, we show that there is no need to reverse-engineer the CAN protocol to identify drivers. In addition, re-identification remains feasible without significantly constraining the driving environment; what's more, such inference is practical with
 off-the-shelf machine learning techniques readily available to anybody. This is a plausible scenario when a malicious actor gains access to the car's CAN bus through an Internet-connected device such as a smartphone using a wired or wireless data collector adapter plugged into the car's OBD-II connector \cite{szalay2015ict}. 
 For the purpose of inference, we blindly consider all possible time series that could be easily extracted from CAN logs without any sort of manual inspection of their content. We use machine learning techniques to select time series with sufficient predictive power and use their combination for classification.  Our solution is scalable and does not require reverse-engineering the proprietary communication protocol used on the CAN bus.

Our specific contributions are twofold:
\begin{itemize}
    \item We demonstrate the feasibility of driver re-identification even if drivers follow different routes and the exact signals (such as velocity, acceleration, RPM, etc.) cannot be extracted directly from the captured CAN log. We extract a time series from every byte of every CAN message, from which the ones with the most predictive power are identified.
    In particular, we build a convolution-based neural network, where features from individual time series are learnt automatically by convolutional layers, which are then combined in a mixture model in order to perform the final classification.  
    We achieve a mean accuracy of driver classification between 75-85\% for CAN traces with a length of less than 2 minutes. By contrast, most prior works \cite{HallacSSLHRSL16, EnevTKK16} achieved comparable result only with fewer drivers (2-15) and when the exact signals could be extracted and are readily available for classification.
    \item We propose a scalable technique to classify a large set of time series even if many of the individual time series are significantly noisy and hence lack sufficient predictive power individually. Our approach first classifies each time series separately, and combines the best performing individual models to classify the whole set of time series. 
\end{itemize}






\section{Background}
\subsection{Signal Extraction from Controller Area Network (CAN)}
\label{sec:can}

The Controller Area Network (CAN) is a bus system providing in-vehicle communications for ECUs and other devices~\cite{szalay2015ict}. 
The overwhelming majority of cars use one or more proprietary CAN protocols. Generally, sensor signals in CAN variants have a sampling frequency in the order of 10 ms$^{-1}$, i.e., an ECU packs the measured signal samples into a CAN message and broadcasts that on the CAN bus every 10 ms.

CAN messages can be eavesdropped through the standard on-board diagnostics (OBD, OBD-II) port. 
Sensor signals carried by OBD have a sampling frequency in the order of 1 second. However, in certain vehicle makes and models, one or more CANs are also connected to the OBD-II port. In such cars (like the one used in this paper), also utilizing OBD over the CAN physical layer, it is possible to extract fine-grained CAN data via an OBD-II logger device~\cite{szalay2015ict}.
Indeed,  collecting in-vehicle network logs is not a privilege of car companies any more. For example, insurance companies sell a complete kit which consists of a smartphone application as well as an ODB-II adapter. This adapter is then voluntarily installed by the driver and wirelessly connected to the driver's smartphone through Bluetooth\footnote{Other malicious parties may have access to the driver's device by installing a malicious application which then may gain access to the adapter.}.  

However, owing to the confidential message format and semantics, extracting a sensor signal from the captured CAN messages is not straightforward. Table \ref{tab:can_msg} shows a simplified picture of a CAN message\footnote{This example shows an already stripped message, i.e., we do not discuss end of frame or check bits.}. A CAN bus message has several components including the timestamp, a message identifier (CAN-ID)
, a Remote Transmission Request which allows ECUs to request messages from other ECUs, and the actual data field with a size of at most 8 bytes. The Data field can be broken to sensor signals which are then transformed and/or converted to a human-readable format in order to enable further analysis~\cite{lestyan2019}. 

For example, the position of the brake pedal may be carried by the fourth byte of the message with ID 0x02c4 in Table \ref{tab:can_msg}. Therefore, extracting the fourth byte of every consecutive instance of this message type (ordered by their timestamp) one can reconstruct the braking signal (i.e., position of the brake pedal over time, which is shown in green in Table \ref{tab:can_msg}).
\emph{Importantly, the message ID, offset and length of a signal are not standardised and kept confidential by OEMs.} Consequently, extracting a signal from recorded CAN messages would require (partly) reverse-engineering the specific CAN protocol which may be illegal under practical circumstances~\cite{bowers_legal}. 

\begin{table}[!tb]
    
\centering
\caption{Example of CAN messages and the extracted time series\label{tab:can_msg}. The red, green, and blue time series are obtained by extracting the 1st, 4th and 7th byte of every time-ordered CAN message with ID 0x02c4, respectively.}
\begin{subtable}{0.97\textwidth}
\scalebox{0.63}
{
\begin{tabular}{|c|l|c|c|l|}

  \hline
  Timestamp & CAN-ID & Req & Len & Data \\
  \hline
 
    \texttt{1481492683.285052} & \texttt{0x0208} & \texttt{000} & \texttt{0x8} & \texttt{0x00 0x00 0x32 0x00 0x0e 0x32 0xfe 0x3c} \\
  \texttt{1481492674.736055} & \texttt{0x02c4} & \texttt{000} & \texttt{0x8} & \texttt{\textcolor{red}{0x82} 0xc8 0x00 \textcolor{ForestGreen}{0x0f} 0x03 0x00 \textcolor{blue}{0x92} 0x3c} \\
  \texttt{1481492674.736055} & \texttt{0x02c4} & \texttt{000} & \texttt{0x8} & \texttt{\textcolor{red}{0x82} 0xc9 0x00 \textcolor{ForestGreen}{0x0f} 0x00 0x00 \textcolor{blue}{0x92} 0x4c} \\
  \texttt{1481492674.736055} & \texttt{0x02c4} & \texttt{000} & \texttt{0x8} & \texttt{\textcolor{red}{0x82} 0xcc 0x00 \textcolor{ForestGreen}{0x0f} 0x08 0x00 \textcolor{blue}{0x92} 0x5a} \\
  \texttt{1497323915.123844} & \texttt{0x018e}  &  \texttt{000}  &  \texttt{0x8} &  \texttt{0x03 0x03 0x00 0x00 0x00 0x00 0x07 0x3f} \\
  \texttt{1497323915.112910} &  \texttt{0x00f1} & \texttt{000} & \texttt{0x6} & \texttt{0x28 0x00 0x00 0x40 0x00 0x00} \\
    \texttt{1481492674.736055} & \texttt{0x02c4} & \texttt{000} & \texttt{0x8} & \texttt{\textcolor{red}{0x82} 0xd2 0x00 \textcolor{ForestGreen}{0x0f} 0x0c 0x00 \textcolor{blue}{0x92} 0x5d} \\
     \texttt{1481492674.736055} & \texttt{0x02c4} & \texttt{000} & \texttt{0x8} & \texttt{\textcolor{red}{0x82} 0xa1 0x00 \textcolor{ForestGreen}{0x0f} 0xa1 0x00 \textcolor{blue}{0x92} 0x4d} \\
  \hline
 
\end{tabular}
}
\end{subtable}

\begin{subtable}{\textwidth}
\vspace{3mm}
  \includegraphics[width=0.49\linewidth]{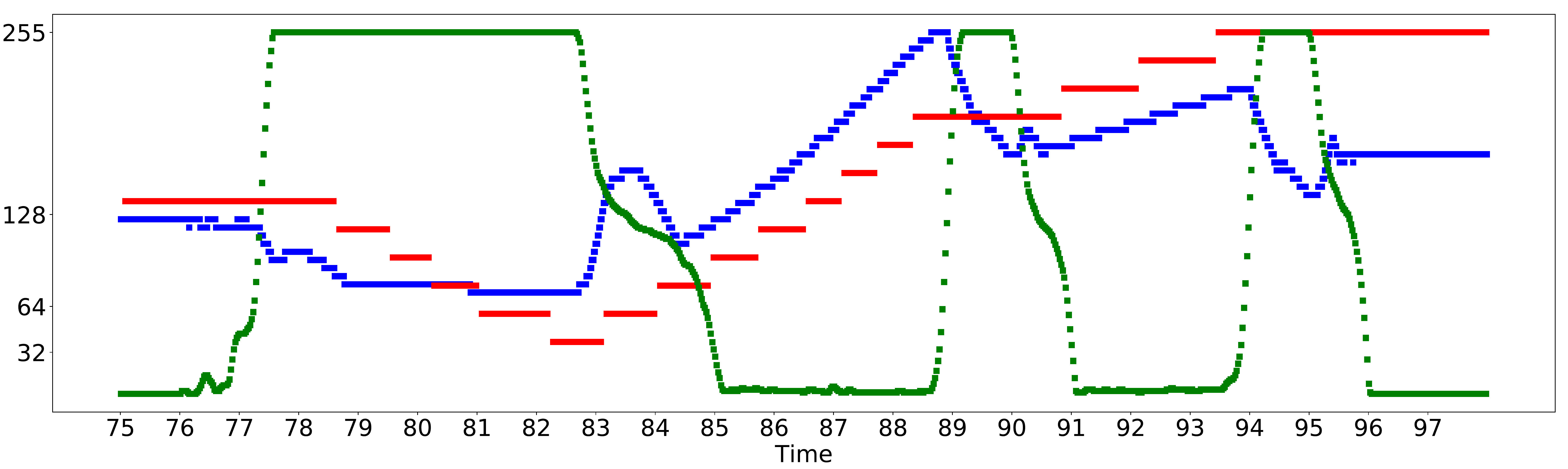}
\end{subtable}
\end{table}

\subsection{Convolutional Neural Networks for Automatic Feature Extraction}
 In  recent  years,  convolutional  neural  networks  (CNNs) have significantly progressed the fields of audio and visual pattern recognition  \cite{CunBDHHHJ89}. CNNs are capable of automatically learning complex feature representations using their convolutional layers. Specifically, CNNs produce a set of (learnable) filters that can recognize patterns in a translation-invariant way.
 Suppose $f$ is a filter (or vector) with size $n$ and $T$ is a time series also represented by a vector of values.
 Then, the 1-dimensional discrete convolution of $f$ and $T$ is given by 
 $$
(T*f)[i] = \sum_{j=1}^{n} f[n+1-j]\cdot T[i+j-1]
 $$
 Filters are learnt in order to extract different features from the time series using convolution. For instance, if $f=[1,-1]$, $T*f$ equals the gradient (i.e., difference) between any two neighboring points. Hence, there is no need to manually craft such features since the value of $T*f$ (or its simple functions such as maximum or average) is used as "features" and are further fed into a classification model. By changing the number and size of filters as well as stacking multiple convolution layers, a CNN can learn to detect many different local traits of a time series. For example, if $T$ is the velocity, then convolving $T$ with $f=[1,-1]$ yields the acceleration. Therefore, by applying max pooling on the convolved signal the model can learn to distinguish drivers based on their maximum acceleration. 
 Although, in theory, CNN could learn many different transformations of the time series such as discrete Fourier transform, such specific functions are almost never learnt in practice \cite{abs-1807-06399}. It is still an open question what exact functions are learnable by CNNs.

If $(T*f)$ is evaluated at every point of $T$, then the same feature represented by $f$ can be detected along the whole time series $T$. This makes CNNs particularly suitable to detect local features. Indeed, most drivers exhibit only local distinguishing features (e.g., the short moments when they accelerate or decelerate) as the signal is globally shaped by traffic conditions rather than by the driver's behaviour. 
 
CNNs are also more scalable than traditional fully connected neural networks (FCNNs), because the same filters are applied along the whole time series, that is, the number of learnable parameters is proportional to the number and size of filters and not to the size $|T|$ of the whole time series. This also makes CNNs less prone to overfitting.

 \subsection{Long Short Term Memory Networks}
 A Recurrent Neural Network (RNN) uses memory to model short-term dependencies in sequential data. When folded out in time, it can be considered as a feedforward deep neural network with indefinitely many layers. 
RNNs generally receive input and produce output at each time step. 
Long-term temporal dependencies are difficult to model with RNNs because the gradient of the loss function decays exponentially with time (the vanishing gradient problem).
Another weakness of common RNNs is that they do not explicitly support multiple time scales, and any temporal hierarchy that is present in the input signal needs to be embedded implicitly in the network. 
However, time series often have a temporal hierarchy, where the information is scattered over multiple time scales. 
Long Short-Term memory (LSTM) networks \cite{hochreiter1997long}, a special type of RNNs, overcome the above problems. The recurrent weights in an LSTM (learnt during the training phase) select which information they need to pass onwards vs. which they need to discard. A common LSTM unit is composed of a cell, an input gate, an output gate and a forget gate. The cell remembers values over arbitrary time intervals, while the three gates regulate the flow of information into and out of the cell.

\section{Inference model}
\label{sec:model}

\subsection{Overview}

 Our premise is that the signals represented by the time series are unknown, therefore, for the purpose of inference, we use an artificial neural network (ANN), where the features of a time series are automatically extracted and learnt by multiple convolution layers. This is advantageous as the manual feature extraction of an unknown signal would be difficult otherwise. 
 
 We propose a mixture model with two main components: 
 \begin{enumerate}
     \item \textbf{Individual time series (ITS) model:}  This model is specialized to the classification of a single time series. We build one ITS model per time series, hence we obtain $N$ different ITS models altogether, where $N$ is the number of all time series. A time series is extracted by concatenating the same byte of every CAN message with identical message ID.
     \item \textbf{Mixture model:} The already trained ITS models are combined into a single fully connected decision/ouput layer which provides the final classification result. 
 \end{enumerate}
Importantly, ITS models are trained individually and separately from the mixture model, which makes our solution scalable. In particular, each ITS model can be trained independently and in parallel, and the training of the whole mixture model does not require to fine-tune or re-train the already trained ITS models. Once ITS models are trained, their weights are frozen and not modified during the training of their mixture model. 

\begin{figure*}[h!]
	\centering
	\includegraphics[width=1\textwidth]{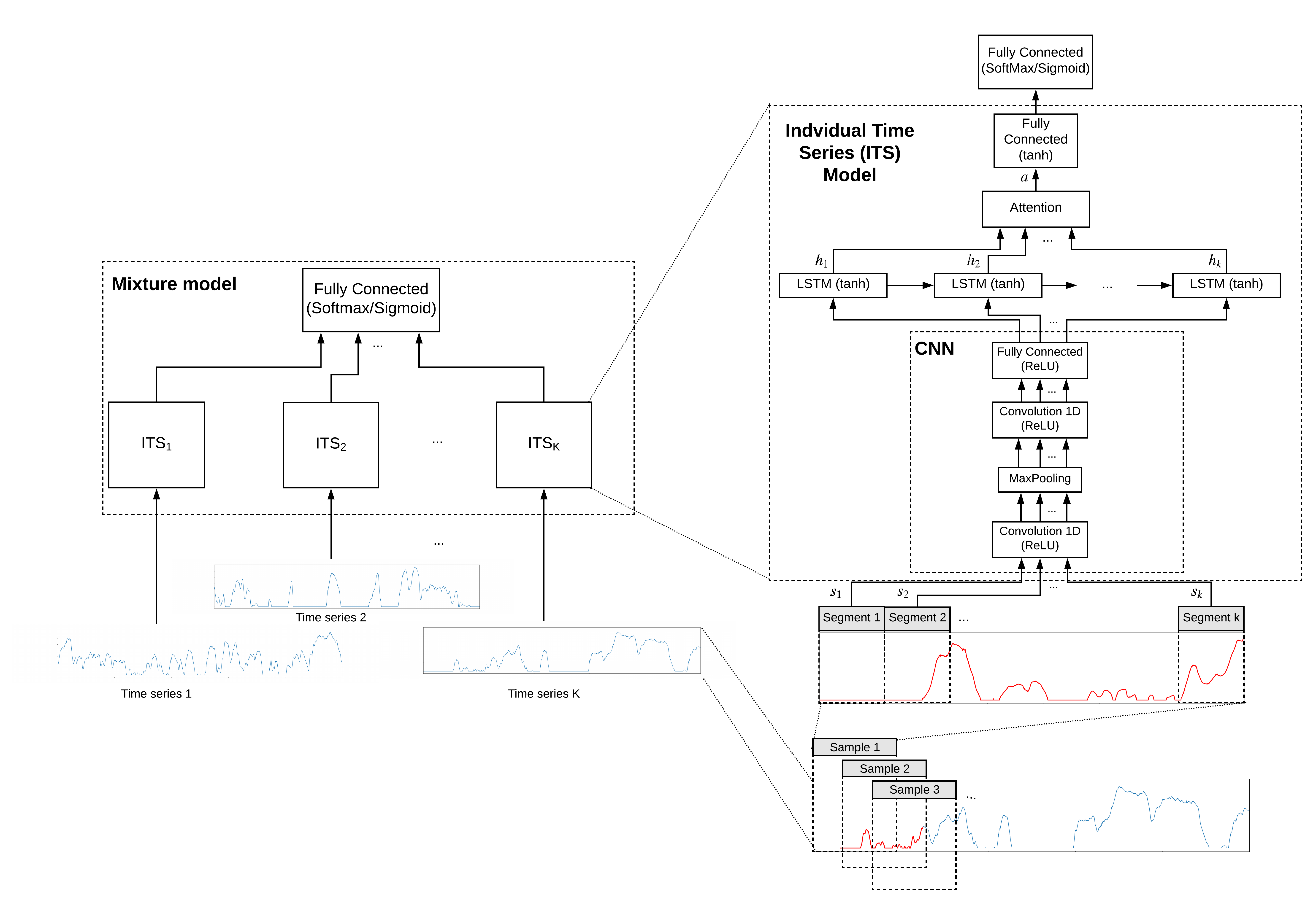}
		\caption{Our mixture model.}
		\label{fig:model}
	
\end{figure*}

Next, we describe the model architecture in more details. The architecture of our proposal is illustrated in Figure \ref{fig:model}.

\subsection{Data augmentation}
\label{sec:data_aug}
We first divide the time series into equally-sized segments, from here on referred to as samples. Given a single sample as input, our model attempts to classify it into its correct class (e.g., driver ID).  Due to the limited amount of  data, and similarly to prior works \cite{DBLP:journals/corr/CuiCC16, YaoHZZA17}, we use a sliding window over the whole time series to create samples. Specifically, let $T = (t_1, t_2, \ldots, t_{|T|})$ denote a time series, and let $T_{i:j} = (t_i, t_{i+1}, \ldots, t_j)$ be a window of $T$ between positions $i$ and $j$ ($1\leq i \leq j \leq |T|$). Then, the set of all $n$ samples created from $T$ is given by $\{T_{1,n}, T_{1+\ell,n+\ell}, \ldots, T_{|T|-\ell+1,|T|}\}$, that is, all consecutive windows shifted by $\ell$ time slots. 


\subsection{Individual Time Series (ITS) Model}

We use a CNN combined with Long Short-Term Memory (LSTM) \cite{hochreiter1997long}. Our ITS model is inspired by~\cite{YaoHZZA17}; however, it also differs  in several aspects as we explain in Section \ref{sec:related}.

First, the input sample is divided into equally-sized \emph{non-overlapping} $k$ segments. 
Each segment $s_i$ of a sample is transformed into a higher-level representation/features $\mathrm{CNN}(s_i)$ using the same CNN model. The CNN model is composed of two 1-dimensional convolutional layers (with Rectified Linear Unit activations), which are separated by a max pooling layer in order to reduce model complexity, and a fully connected layer producing the output of the CNN. 
The segment size should be sufficiently large in order to capture any human driving reaction (e.g., acceleration, deceleration, gear switching, etc.), but still small enough to focus on distinctive driver features.  

As the ITS model is applied on each segment of the sample independently, it can only model local features of a segment except for the sequential information along the sequence of segments as well as any global features of the whole sample. For this reason, we use an LSTM layer (with $\mathsf{tanh}$ activation) and it is applied on the sequence of local features extracted by CNN.
In particular, a single CNN model is used to extract features from every segment $s_i$. The sequence of these representations are further processed by  LSTM, which computes time-dependent features by transforming the input sequence into another sequence $h_1, h_2, \ldots, h_k$ composed of the hidden states of the LSTM model. Specifically, for input segment $s_i$, $h_i = \mathrm{LSTM}(\mathrm{CNN}(s_i), h_{i-1})$.
A hidden state $h_i$ is produced for every input segment $s_i$, dependent on the current input as well as all previous input segments through $h_{i-1}$. The sequence of $h_1, h_2, \ldots, h_k$ is fed into an attention layer in order to select those parts of the sequence which represent distinctive driver features. 
In particular, a sample may contain only a few segments with clear distinctive features of the drivers such as a unique pattern of speeding, steering, or braking, while the rest may lack any unique driving pattern and hence bearing less importance in distinguishing drivers. 

LSTM is ineffective to model sequential dependencies if the input sequence is too long (i.e., there are too many segments), which is alleviated by attention as follows. 
Several attention mechanisms have been  proposed in the literature mainly for text classification and machine translation \cite{BahdanauCSBB16, RaffelE15}. In this work, we use a simple mechanism \cite{RaffelE15} which provides the best performance for our dataset. The attention mechanism computes the weighted average of all the hidden states $h_1, h_2, \ldots, h_k$ produced by LSTM along the whole input sequence. 
More specifically, the attention mechanism is defined as
\begin{align*}
u_{i} &= \mathrm{tanh}(w^{\top} h_{i} + b) \\
\alpha_{i} &=  \frac{\exp(u_{i})}{\sum_i\exp(u_{i})} \\
a &= \sum_i \alpha_{i} h_{i} 
\end{align*}
where $w \in \mathbb{R}^{k}$ and $b \in \mathbb{R}$ are jointly learnt parameters, and $a_i$ is the attention vector which represents sample $i$. In essence, $a$ is the weighted average of each segment's LSTM representation, where the weights are a non-linear function of these representations. This function is modeled by a shallow neural network with weights $w$ and bias $b$. 

Finally, the attention vector $a$ is provided to a fully connected layer with $\mathsf{tanh}$ activation, whose output is then given to a final softmax (multi-class classification) or a sigmoid layer (binary classification). 
We apply dropout regularization on every fully connected layer of our model. 


\subsection{Mixture model}
ITS models classify individual time series and therefore each can have very different predictive performance. Indeed, some time series contain more distinctive driver features (e.g., brake pedal position) than others (e.g., velocity). In fact, it turns out that the majority of time series often do not have any predictive power at all.

We combine ITS models into a single Mixture model by connecting all ITS models to the same fully-connected neural layer, called mixture layer, whose output provides the final prediction. We first remove the output layer of the already trained ITS models, and merge the remaining part into the mixture layer.  Finally, the whole mixture model is trained by adjusting the weights of only the mixture layer while keeping all parameters of the individual models intact.
In case there are too many ITS models and the training of the mixture model is not feasible due to memory constraints, only the top-K best performing ITS models are used in the mixture. Also, the trained ITS models can be compressed by distillation \cite{HintonVD15}. 

In fact, our mixture model is a sort of committee machine \cite{haykin99a}, 
where multiple neural networks, called as experts, are combined to provide the final classification. Indeed, each ITS model is an expert which is specialized to extract a high
-level representation of a single time-series. These representations are non-linearly combined to produce the classification.

The benefit of the mixture layer is to model inter-dependencies between different time series which may improve classification accuracy. For example, the acceleration and deceleration patterns individually may not be unique to a driver but their combination is.


\section{Evaluation}

In this section, we evaluate our mixture model described in Section \ref{sec:model}.
 

\subsection{Dataset}
\label{sec:data}

As CAN data logs are not widely available, we conducted a measurement campaign. For data collection in particular we connected a logging device to the OBD-II port and logged all observed messages from various ECUs. Such a device acts as a node on the CAN bus and is able to read and store all broadcast messages. Our team developed both the logging device (based on a Raspberry PI 3) and the logging software (in C).\footnote{Note that it is common that the OBD-II connector is found under the steering wheel. Also note that not all car makes and models connect the CAN serving the drive-train ECUs (or any CAN) to the OBD-II port (e.g., Volkswagen, BMW, etc.).}


Most routes were driven inside or close to Budapest; approximately 15-20\% was recorded on a motorway.
Drivers were free to choose their way, but still conforming to three practical requirements: (1) record at least 15 minutes of driving in total, (2) do not record data when driving up and down hills, (3) do not record data in extremely heavy traffic (short runs and idling). Free driving was recorded for all 33 drivers with an Opel Astra 2018: 11 people were between the age of 20-30, 8 of 25-30, 7 of 35-40 and and 7 above 40; there were 5 women and 28 men; 12 with less experience (less than 7000 km per year on average or novice driver), 11 with average experience (8-20000 km per year), and 10 with above average experience (more than 20000 km per year). 
All drivers drove between 10am and 4pm on weekdays and the average trace length is 29.81 minutes with a standard deviation of 13.48, the shortest trace is 17.71 minutes long, whereas the longest trace is 85 minutes long.
The data were collected with the users’ explicit consent after informing them about the purpose of the study. Exact identity of drivers was not recorded except for their personal attributes described above. Traces of different drivers were linked together using random identifiers.

\begin{table}[h!]
	\centering
	\caption{Dataset summary.}
	\begin{tabular}{|c|c|c|}
	    \hline
		\emph{Attribute} & \emph{Value} & \emph{Population} \\
		\hline
		\hline
		\multirow{2}{*}{Gender} & Male &  28 \\
		& Female & 5  \\
		\hline
			\multirow{4}{*}{Age} & [20-25] & 11  \\
		& [25-30] & 8   \\
		& [30-40]  & 7   \\
		& [40-70]  & 7   \\
		\hline
		\multirow{3}{*}{Experience} & Low & 12   \\
		& Average & 11   \\
		& High  & 10   \\
		\hline
	\end{tabular}
	
	\label{tab:dataset}
\end{table}

\subsection{Extracting time series from CAN logs}
The CAN logs of all drivers contain 53 different message IDs from which we extract 72 different time series per driver.  More specifically, we  consider all single bytes of every message,  and a single time series corresponds to the series of bytes retrieved from the same byte position of all consecutive messages with the same message ID, which is also illustrated in Table \ref{tab:can_msg}. 
We dropped all time series (1) with a single constant value in all time slots, (2) that were too short, (3) contained the values of counters.
We retained the time series which occur in each driver's trace. This pre-processing finally yields 72 time series per driver.
Accordingly, we built 72 different ITS models, each corresponding to a single time series. The validation accuracy of each ITS model is evaluated, and the top-10 best performing ITS models are selected and used in the mixture model.  

\subsection{Training and Testing data}
In order to train an ITS model which is specialized for the classification of a single time series, we build two datasets $\mathbb{T}$ and $\mathbb{S}$, which are further used to create the training and testing datasets for any classification problem as follows.

First, each time series of every driver is divided into two parts. Samples extracted from the first part of every trace, using the sliding window technique described in Section \ref{sec:data_aug}, are added to $\mathbb{T}$. $\mathbb{S}$ is created analogously from the second part of every trace. Since the size of the first part is always set to 80\% of the whole trace, the training samples will roughly cover 80\% of all the samples from all drivers. Training data is obtained by picking an identical number of samples per class from $\mathbb{T}$, and testing data is generated from $\mathbb{S}$.
These guarantee that (1) the training and testing sets do not overlap, (2) training data is always balanced, and (3) the proportion of samples from a driver is identical both in the training and the testing data (i.e., there is no dataset shift).

\subsection{Hyperparameters}
Sample sizes are varied among 20, 60, and 120 seconds, and consecutive samples are shifted by 0.1 second. For ITS models, samples are divided into non-overlapping segments with a size of 3 seconds, which is believed to be sufficient to capture the most important driving actions. Each convolutional layer has a kernel size corresponding to 0.5 seconds with a stride of 0.05 seconds. The first convolution layer has 20 filters while the second layer applied on the output of the first layer has 40 filters followed by batch normalization \cite{IoffeS15}. Max pooling is applied on the result of every 5 consecutive convolution operations of the first layer. All FCNNs are composed of 64 units, and the number of LSTM's hidden states is 16. ITS models are trained as long as validation accuracy improved: 2 epochs for ITS models and 4 epochs for the Mixture model on average.  All models are trained with RMSprop\footnote{\url{www.cs.toronto.edu/~tijmen/csc321/slides/lecture_slides_lec6.pdf}} with a fixed learning rate of 0.001.
The output layers use softmax activation for multi-class classification and sigmoid activation for binary classification, which are optimized against categorical and binary cross-entropy, respectively, for all models.

\subsection{Results}
\label{sec:results}
We use the model described in Section \ref{sec:model} for all classification tasks\footnote{We implemented our model in Python 3.6.7 using the framework of Keras 2.2.2.
Evaluation was done on commodity hardware with Intel Core i7-4770 CPU and 32 GB RAM, where the average time training of a mixture model is 294 sec, and one ITS is trained in 157 sec on average. A powerful GPU can considerably speed up training.}.
To measure performance on the testing data, we define testing accuracy as all correct classifications (TP$+$TN, True Positive plus True Negative) divided by all classifications:
\begin{equation*}
    \mathrm{Acc} = \frac{\mathrm{TP} + \mathrm{TN}}{\mathrm{All}}
\end{equation*}


\subsubsection{1-vs-all}
Table \ref{tab:reid} shows the re-identification accuracy in the 1-vs-all scenario, i.e., when one driver is distinguished from all other drivers with a single binary mixture model. We show the mean value over all 33 drivers (i.e., the mean accuracy over all the 33 models), standard deviation, and the extreme values for a specific driver. It is easy to see that accuracy improves sublinearly with the length of the sample trace, the mean ranging from 75\% up to 85\%. Maximum accuracy values for the most correctly classified driver are between 95\% and 100\%, while minimum values for the hardest-to-classify driver are between 43\% and 46\%.

Table~\ref{tab:dataset} depicts the 1-vs-all re-identification accuracy depending on the personal attribute values, which include gender, age, and driving experience.
Interestingly, the most experienced drivers are the easiest to distinguish from all the other drivers with a mean accuracy of 86\%, which shows that, by practicing, people tend to develop more unique skill sets. This phenomenon has also been confirmed in other domains \cite{IslamHLNVYG15}. 

\begin{table}[tb]
\centering
\caption{Re-identification accuracy: 1-vs-all}
\begin{tabular}{|lrrrr|}
\hline
\emph{Sample length} &      \emph{Mean} &      \emph{Std.dev.} &       \emph{Max} &       \emph{Min} \\
\hline \hline
20~s  &  0.758 &  0.017 &  0.949 &  0.432 \\
60~s &  0.829 &  0.019 &  0.995 &  0.436 \\
120~s &  0.847 &  0.025 &  1.000 &  0.465 \\
\hline
\end{tabular}
\label{tab:reid}
\end{table}

\begin{table}[tb]
	\centering
	\caption{Re-id. accuracy: 1-vs-all, sample length: 60 secs}
	\begin{tabular}{|c|c|c|c|c|c|c|}
	    \hline
		\emph{Attribute} & \emph{Value} & \emph{Persons} & \emph{Mean} & \emph{Std} & \emph{Max} & \emph{Min} \\
		\hline
		\hline
		\multirow{2}{*}{Gender} & Male &  28 & 0.814 &  0.140 &  0.995 &  0.436\\
		& Female & 5  & 0.911 &  0.099 &  0.968 &  0.734 \\
		\hline
			\multirow{4}{*}{Age} & [20-25] & 11 & 0.840 &  0.106 &  0.939 &  0.573 \\
		& [25-30] & 8   &  0.788 &  0.184 &  0.973 &  0.436\\
		& [30-40]  & 7  &  0.892 &  0.143 &  0.995 &  0.606 \\
		& [40-70]  & 7 &  0.806 &  0.126 &  0.969 &  0.586   \\
		\hline
		\multirow{3}{*}{Experience} & Low & 12 & 0.834 &  0.108 &  0.968 &  0.573  \\
		& Average & 11  & 0.799 &  0.172 &  0.964 &  0.436  \\
		& High  & 10  & 0.860 &  0.135 &  0.995 &  0.586\\
		\hline
	\end{tabular}
	\label{tab:dataset}
\end{table}

\subsubsection{Many-vs-all}
\begin{figure}[htb]
\centering
\includegraphics[width=0.45\textwidth]{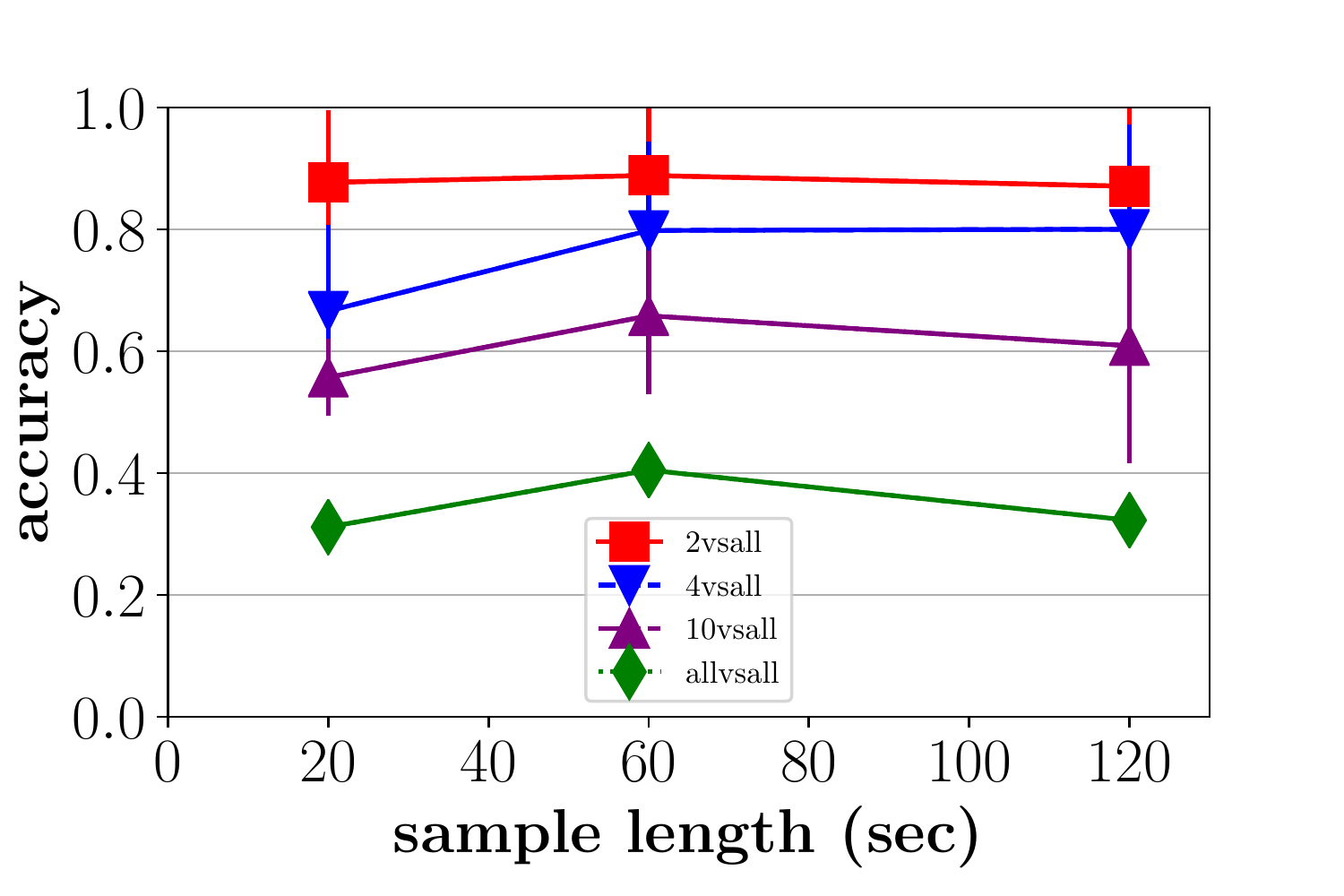}
\caption{Re-identification accuracy: many-vs-all}
\label{fig:reid}
\end{figure}

Fig. \ref{fig:reid} shows the re-identification accuracy in the many-vs-all scenario, i.e., when a group of drivers is distinguished from all other drivers with a single multi-class mixture model, further characterizing the performance of our method. In the 2-vs-all case, we randomly select 50 different groups of 2 drivers out of the possible ${33 \choose 2}$, and plot the mean value and the standard deviation; we do similarly with group sizes of 4, 10 and 33 (with only 1 possible allocation in the last, all-vs-all case). Notice that all-vs-all corresponds to the case when each driver is distinguished from each other using a single model. We observe that the mean value of accuracy is higher when the group size is smaller, ranging from the highest mean of 40\% (all-vs-all, trace length 60~s) up to 91\% (2-vs-all, trace length 20~s). It is interesting that a larger sample length does not always translate into higher accuracy; in fact, in almost all displayed cases the highest mean accuracy comes from models built on 60 s long samples (the exception to the rule is the 4-vs-all case, where the 120 s sample length yields slightly better results). Note that even in the hardest, all-vs-all case, our approach provides a 13-fold accuracy improvement compared to the $1/33 = 3.03\%$ baseline.

\section{Related work}
\label{sec:related}
In this section, we give an overview on the related work with regard to driver re-identification and neural network based time series classification.

\subsection{Driver re-identification}
\label{sec:driver_related}

Driver characterization based on driving data has gathered significant research interest from both the automotive control and the data privacy domain. In automotive control, substantial research efforts were made in modeling and recognizing driver behavior based on CAN and other driving data with the objective of designing and improving intelligent vehicle systems and providing enhanced driver safety features~ \cite{wang2014modeling, lin2014overview}. On the other hand, the potential re-identification of drivers (i.e., singling out) based on recorded CAN data also became a focus for data privacy research~\cite{EnevTKK16}. The common trait in these works is the presumed familiarity with the specific CAN protocol stack including the presentation and application layers giving the researchers access to sensor signals. This knowledge is usually gained via access to the OEM's documentations in the framework of some research cooperation. As such, researchers do not disclose such information to honor their contract and preserve confidentiality.


Miyajima et al. collected and investigated sensor signals in both a driving simulator and a real vehicle. Using car-following patterns and spectral features of pedal operation signals authors achieved an identification rate of 89.6\% for the simulator (12 drivers) and 76.8\% (276 drivers) for the field test.
Hallac et al. \cite{hallac2016driver} discovered that maneuvers during turning could fingerprint drivers using a dataset provided by Audi. Via the same dataset, Fugiglando et al. \cite{driving_dna} showed that four behavioral traits, namely braking, turning, speeding and fuel efficiency could characterize a driver adequately well. 
Zhang et al. \cite{Zhang:2016:DCB:2856767.2856806} developed a driver classification model based on both CAN and mobile phone sensor data. While effective for distinguishing between pairs of drivers, authors claim that their method does not scale well.
Enev et al. \cite{EnevTKK16}  make use of mostly statistical features as an input for binary (one-vs-one) classification with regard to driving behavior. These binary classifiers are combined into a multi-class classifier, which is not a scalable approach if there is a large number of drivers.
Driving the same car in a constrained parking lot setting and a longer but fixed route, authors re-identified their 15 drivers with 100\% accuracy. Authors had access to all available sensor signals and their scaling and offset parameters from  the manufacturer's documentation. 
Fugiglando et al.~\cite{drivers_classification} developed a new methodology for near-real-time classification of driver behavior in an uncontrolled setting. Despite their advanced use of unsupervised machine learning techniques they conclude that clustering drivers based on their behavior remains a challenging problem.


\subsection{Time series classification with neural networks}
\label{sec:tsc_related}
The field of time series classification (TSC) has proposed a multitude of algorithms. These algorithms are either distance-based, using k-NN classifiers over distance measures between time series; or feature-based, extracting deterministic features in either the time domain or spectrum and applying traditional classification algorithms. Lately, ensemble methods using multiple classifiers have also been studied. An overview and comparative performance evaluation of these methods are presented in \cite{Bagnall2017}. Using neural networks is a novel direction in TSC \cite{DBLP:journals/corr/CuiCC16,WangYO17, resnet, fawaz2018}. In order to perform adequately, these methods require either additional pre-processing of the input signals or very large datasets. For a comparative performance evaluation we refer the reader to the work of Fawaz et al.~\cite{fawaz2018}. 




Our proposed classifier is most similar to the DeepSense framework \cite{YaoHZZA17} and the Multi-channels Deep Convolutional Neural Networks (MC-DCNN) \cite{zheng2014time}. Both works use CNNs to extract features from multiple time series which are fed into a Multi-Layer Perceptron (MLP) \cite{zheng2014time} or a Gated Recurrent Unit (GRU) \cite{YaoHZZA17}. Our mixture model can be considered as a synergy of these two approaches. Unlike DeepSense \cite{YaoHZZA17} which applies 2 dimensional convolution on multiple time series, we apply 1 dimensional convolution on individual time series which are then combined in a fully connected layer as in MC-DCNN. We found this approach to be superior especially if many of the input time series are noisy and have no predictive power at all. Our approach can also be used to localize the parts of CAN messages which indeed identify a driver.
Our model is also simpler (has fewer parameters) and hence is less prone to overfitting. Unlike MC-DCNN, we use LSTM with an attention layer per time series to focus on its most distinctive parts. 

\section{Conclusion and Future Work}
Our study demonstrates that driver re-identification remains feasible even without the manual inspection of messages captured on the CAN bus. In other words, profiling drivers can be performed even without reverse-engineering the CAN protocol.

The empirical evaluation of our approach on a dataset of 33 drivers demonstrates that a driver can be recognized with a success probability of 85\% based on a mere 2 minutes of driving in urban areas, even if drivers follow different routes. Our findings include that drivers with substantial experience are the easiest to distinguish.
As features are extracted automatically using convolutional neural networks, we conjecture that our technique is also applicable to predicting various other driver attributes such as age, gender, or driving experience; however, the necessary empirical justification would require a significantly larger dataset. 

Our study does not only raise the flag to car drivers, but also to companies collecting in-vehicle network logs; the re-identification (and/or profiling) of drivers so effortlessly means that CAN logs indeed constitute personal data and, as such, are subject to the European General Data Protection Regulation (GDPR) \cite{GDPR} as of 25 May 2018. 
It remains an open problem whether more sensitive attributes such as health status 
are also predictable from CAN logs. 

Our work also advocates the widespread adoption of standardized security protocols for the protection of CAN network traffic \cite{GrozaMHV17}. Although ad hoc solutions like dispersing the bits of a sensor signal within a CAN message would make our approach less effective, we conjecture that extracting signals from such obfuscated CAN messages remains feasible with appropriate statistical approaches.

It also remains an open problem whether captured CAN logs can be effectively anonymized.  Although car companies stress that the collected network logs are ``anonymized'', the proper anonymization of such high-dimensional data is notoriously difficult in practice without significantly degrading accuracy  \cite{aggarwal2005}. We believe that most viable approaches include the release of aggregate noisy statistics with provable  privacy guarantees \cite{Dwork06}.

\section{Acknowledgement}
Gergely Acs was supported by the Premium Post Doctorate Research Grant of the Hungarian Academy of Sciences (MTA). This work has been partially funded by the Eu- ropean Social Fund via the project EFOP-3.6.2-16-2017-00002, by the European Commission via the H2020-ECSEL-2017 project SECREDAS (Grant Agreement no. 783119) and the Higher Education Excellence Program of the Ministry of Human Capacities in the frame of Artificial Intelligence re- search area of Budapest University of Technology and Economics (BME FIKP-MI/SC).

\bibliographystyle{IEEEtran}
\small
\bibliography{main}

\begin{thebibliography}{10}
\providecommand{\url}[1]{#1}
\csname url@samestyle\endcsname
\providecommand{\newblock}{\relax}
\providecommand{\bibinfo}[2]{#2}
\providecommand{\BIBentrySTDinterwordspacing}{\spaceskip=0pt\relax}
\providecommand{\BIBentryALTinterwordstretchfactor}{4}
\providecommand{\BIBentryALTinterwordspacing}{\spaceskip=\fontdimen2\font plus
\BIBentryALTinterwordstretchfactor\fontdimen3\font minus
  \fontdimen4\font\relax}
\providecommand{\BIBforeignlanguage}[2]{{%
\expandafter\ifx\csname l@#1\endcsname\relax
\typeout{** WARNING: IEEEtran.bst: No hyphenation pattern has been}%
\typeout{** loaded for the language `#1'. Using the pattern for}%
\typeout{** the default language instead.}%
\else
\language=\csname l@#1\endcsname
\fi
#2}}
\providecommand{\BIBdecl}{\relax}
\BIBdecl

\bibitem{szalay2015ict}
Z.~Szalay, Z.~K{\'a}nya, L.~Lengyel, P.~Ekler, T.~Ujj, T.~Balogh, and
  H.~Charaf, ``Ict in road vehicles—reliable vehicle sensor information from
  obd versus can,'' in \emph{IEEE MT-ITS}, 2015.

\bibitem{EnevTKK16}
M.~Enev, A.~Takakuwa, K.~Koscher, and T.~Kohno, ``Automobile driver
  fingerprinting,'' \emph{PoPETs}, vol. 2016, no.~1, pp. 34--50, 2016.

\bibitem{HallacSSLHRSL16}
D.~Hallac, A.~Sharang, R.~Stahlmann, A.~Lamprecht, M.~Huber, M.~Roehder,
  R.~Sosic, and J.~Leskovec, ``Driver identification using automobile sensor
  data from a single turn,'' in \emph{IEEE ITSC}.

\bibitem{GMM_driving}
C.~Miyajima, Y.~Nishiwaki, K.~Ozawa, T.~Wakita, K.~Itou, K.~Takeda, and
  F.~Itakura, ``Driver modeling based on driving behavior and its evaluation in
  driver identification,'' \emph{Proceedings of the IEEE}, vol.~95, no.~2, pp.
  427--437, 2007.

\bibitem{Zhang:2016:DCB:2856767.2856806}
C.~Zhang, M.~Patel, S.~Buthpitiya, K.~Lyons, B.~Harrison, and G.~D. Abowd,
  ``Driver classification based on driving behaviors,'' in \emph{IUI}, 2016.

\bibitem{bowers_legal}
S.~Bressman, ``{Restricting Reverse Engineering Through Shrink-Wrap Licenses:
  Bowers v. {Baystate} {Technologies}, Inc.}'' \emph{BUJ Sci. \& Tech. L.},
  vol.~9, p. 185, 2003.

\bibitem{lestyan2019}
S.~Lestyan, G.~Acs, G.~Bicz{\'o}k, and Z.~Szalay, ``Extracting vehicle sensor
  signals from can logs for driver re-identification,'' in \emph{5th
  International Conference on Information Security and Privacy (ICISSP
  2019)}.\hskip 1em plus 0.5em minus 0.4em\relax SCITEPRESS, 2019, shortlisted
  for Best Student Paper Award.

\bibitem{CunBDHHHJ89}
Y.~LeCun, B.~E. Boser, J.~S. Denker, D.~Henderson, R.~E. Howard, W.~E. Hubbard,
  and L.~D. Jackel, ``Handwritten digit recognition with a back-propagation
  network,'' in \emph{NIPS}, 1989, pp. 396--404.

\bibitem{abs-1807-06399}
\BIBentryALTinterwordspacing
M.~Nye and A.~Saxe, ``Are efficient deep representations learnable?''
  \emph{CoRR}, vol. abs/1807.06399, 2018. [Online]. Available:
  \url{http://arxiv.org/abs/1807.06399}
\BIBentrySTDinterwordspacing

\bibitem{hochreiter1997long}
S.~Hochreiter and J.~Schmidhuber, ``Long short-term memory,'' \emph{Neural
  computation}, vol.~9, no.~8, pp. 1735--1780, 1997.

\bibitem{DBLP:journals/corr/CuiCC16}
\BIBentryALTinterwordspacing
Z.~Cui, W.~Chen, and Y.~Chen, ``Multi-scale convolutional neural networks for
  time series classification,'' \emph{CoRR}, vol. abs/1603.06995, 2016.
  [Online]. Available: \url{http://arxiv.org/abs/1603.06995}
\BIBentrySTDinterwordspacing

\bibitem{YaoHZZA17}
S.~Yao, S.~Hu, Y.~Zhao, A.~Zhang, and T.~F. Abdelzaher, ``Deepsense: {A}
  unified deep learning framework for time-series mobile sensing data
  processing,'' in \emph{WWW}, 2017.

\bibitem{BahdanauCSBB16}
D.~Bahdanau, J.~Chorowski, D.~Serdyuk, P.~Brakel, and Y.~Bengio, ``End-to-end
  attention-based large vocabulary speech recognition,'' in \emph{IEEE ICASSP},
  2016, pp. 4945--4949.

\bibitem{RaffelE15}
C.~Raffel and D.~P.~W. Ellis, ``Feed-forward networks with attention can solve
  some long-term memory problems,'' \emph{CoRR}, vol. abs/1512.08756, 2015.

\bibitem{HintonVD15}
\BIBentryALTinterwordspacing
G.~E. Hinton, O.~Vinyals, and J.~Dean, ``Distilling the knowledge in a neural
  network,'' \emph{CoRR}, vol. abs/1503.02531, 2015. [Online]. Available:
  \url{http://arxiv.org/abs/1503.02531}
\BIBentrySTDinterwordspacing

\bibitem{haykin99a}
S.~Haykin, \emph{Neural Networks: A Comprehensive Foundation}.\hskip 1em plus
  0.5em minus 0.4em\relax Prentice Hall, 1999.

\bibitem{IoffeS15}
S.~Ioffe and C.~Szegedy, ``Batch normalization: Accelerating deep network
  training by reducing internal covariate shift,'' in \emph{ICML}, 2015.

\bibitem{IslamHLNVYG15}
A.~C. Islam, R.~E. Harang, A.~Liu, A.~Narayanan, C.~R. Voss, F.~Yamaguchi, and
  R.~Greenstadt, ``De-anonymizing programmers via code stylometry,'' in
  \emph{{USENIX} Security}, 2015.

\bibitem{wang2014modeling}
W.~Wang, J.~Xi, and H.~Chen, ``Modeling and recognizing driver behavior based
  on driving data: A survey,'' \emph{Mathematical Problems in Engineering},
  vol. 2014, 2014.

\bibitem{lin2014overview}
N.~Lin, C.~Zong, M.~Tomizuka, P.~Song, Z.~Zhang, and G.~Li, ``An overview on
  study of identification of driver behavior characteristics for automotive
  control,'' \emph{Mathematical Problems in Engineering}, vol. 2014, 2014.

\bibitem{hallac2016driver}
D.~Hallac, A.~Sharang, R.~Stahlmann, A.~Lamprecht, M.~Huber, M.~Roehder,
  J.~Leskovec \emph{et~al.}, ``Driver identification using automobile sensor
  data from a single turn,'' in \emph{IEEE ITSC}, 2016, pp. 953--958.

\bibitem{driving_dna}
U.~Fugiglando, P.~Santi, S.~Milardo, K.~Abida, and C.~Ratti, ``Characterizing
  the driver dna through can bus data analysis,'' in \emph{ACM International
  Workshop on Smart, Autonomous, and Connected Vehicular Systems and Services},
  2017.

\bibitem{drivers_classification}
U.~Fugiglando, E.~Massaro, P.~Santi, S.~Milardo, K.~Abida, R.~Stahlmann,
  F.~Netter, and C.~Ratti, ``Driving behavior analysis through can bus data in
  an uncontrolled environment,'' \emph{IEEE Transactions on Intelligent
  Transportation Systems}, no.~99, 2018.

\bibitem{Bagnall2017}
\BIBentryALTinterwordspacing
A.~Bagnall, J.~Lines, A.~Bostrom, J.~Large, and E.~Keogh, ``The great time
  series classification bake off: a review and experimental evaluation of
  recent algorithmic advances,'' \emph{Data Mining and Knowledge Discovery},
  vol.~31, no.~3, pp. 606--660, May 2017. [Online]. Available:
  \url{https://doi.org/10.1007/s10618-016-0483-9}
\BIBentrySTDinterwordspacing

\bibitem{WangYO17}
Z.~Wang, W.~Yan, and T.~Oates, ``Time series classification from scratch with
  deep neural networks: {A} strong baseline,'' in \emph{IJCNN}, 2017.

\bibitem{resnet}
K.~He, X.~Zhang, S.~Ren, and J.~Sun, ``Deep residual learning for image
  recognition,'' in \emph{IEEE CVPR}, 2016.

\bibitem{fawaz2018}
H.~I. Fawaz, G.~Forestier, J.~Weber, L.~Idoumghar, and P.-A. Muller, ``Deep
  learning for time series classification: a review,'' \emph{CoRR}, vol.
  abs/1809.04356, 2018.

\bibitem{zheng2014time}
Y.~Zheng, Q.~Liu, E.~Chen, Y.~Ge, and J.~L. Zhao, ``Time series classification
  using multi-channels deep convolutional neural networks,'' in \emph{WAIM},
  2014.

\bibitem{GDPR}
{European Commission}, ``{General European Data Protection Regulation
  (GDPR)},''
  \url{http://eur-lex.europa.eu/LexUriServ/LexUriServ.do?uri=COM:2012:0011:FIN:EN:PDF},
  2012.

\bibitem{GrozaMHV17}
B.~Groza, P.~Murvay, A.~V. Herrewege, and I.~Verbauwhede, ``Libra-can:
  Lightweight broadcast authentication for controller area networks,''
  \emph{{ACM} Trans. Embedded Comput. Syst.}, vol.~16, no.~3, pp. 90:1--90:28,
  2017.

\bibitem{aggarwal2005}
C.~C. Aggarwal, ``On k-anonymity and the curse of dimensionality,'' in
  \emph{VLDB}, 2005, pp. 901--909.

\bibitem{Dwork06}
C.~Dwork, ``Differential privacy,'' in \emph{ICALP}, 2006, pp. 1--12.

\end{thebibliography}

%
%

\end{document}